\documentclass[twocolumn,showpacs,preprintnumbers,amsmath,amssymb]{revtex4}

\usepackage{graphicx}
\usepackage{dcolumn}
\usepackage{bm}

\begin{document}

\title{Spin-polarized transport through a single-level quantum dot in the Kondo regime}
\author{R. \'Swirkowicz$^{1}$}
\author{M. Wilczy\'nski$^{1}$}
\author{ J. Barna\'s$^{2}$}
\email{barnas@main.amu.edu.pl} \affiliation{$^1$Faculty of
Physics, Warsaw University of Technology,
ul. Koszykowa 75, 00-662 Warszawa, Poland \\
$^2$Department of Physics, Adam Mickiewicz University, ul.
Umultowska 85, 61-614 Pozna\'n, Poland;  \\ and Institute of
Molecular Physics, Polish Academy of Sciences, \\ ul.
Smoluchowskiego 17, 60-179 Pozna\'n, Poland}

\date{\today}

\begin{abstract}
Nonequilibrium electronic transport through a quantum dot coupled
to ferromagnetic leads (electrodes) is studied theoretically by
the nonequilibrium Green function technique. The system is
described by the Anderson model with arbitrary correlation
parameter $U$. Exchange interaction between the dot and
ferromagnetic electrodes is taken into account {\it via} an
effective molecular field. The following situations are analyzed
numerically: (i) the dot is symmetrically coupled to two
ferromagnetic leads, (ii) one of the two ferromagnetic leads is
half-metallic with almost total spin polarization of electron
states at the Fermi level, and (iii) one of the two electrodes is
nonmagnetic whereas the other one is ferromagnetic. Generally, the
Kondo peak in the density of states (DOS) becomes spin-split when
the total exchange field acting on the dot is nonzero. The
spin-splitting of the Kondo peak in DOS leads to splitting and
suppression of the corresponding zero bias anomaly in the
differential conductance.
\end{abstract}


\maketitle

\section{Introduction}

The Kondo phenomenon in electronic transport through artificial
quantum dots (QDs) or single molecules attached to nonmagnetic
leads was predicted theoretically more than a decade ago
\cite{glazman88}. Owing to recent progress in nanotechnology, the
phenomenon has been also observed experimentally
\cite{cronenwett98,gores00}. Several theoretical techniques have
been developed to describe this effect
\cite{hershfield91,meir93,yeyati93,wingreen94,kang95,palacios97,scholler00,krawiec02}.
The description is usually simpler in the linear response regime,
where equilibrium methods can be applied, but it becomes more
complex when the system is driven out of equilibrium by an
external bias voltage
\cite{costi00,moore00,wingreen94,krawiec02,kaminski00,lee01}. One
of the methods used to describe non-equilibrium Kondo effect is
the non-equilibrium Green function technique
\cite{meir93,czycholl85,haug_book}. To calculate density of states
(DOS) and electric current  one then needs the retarded/advanced
as well as the lesser (correlation) Green functions. These can be
derived within some approximation schemes.

It has been shown recently that the Kondo effect can also occur
when replacing nonmagnetic leads by ferromagnetic ones
\cite{sergueev02,martinek03a,lu02,bulka03,lopez03}, but
ferromagnetism of the electrodes generally suppresses the effect
-- either partially or totally \cite{martinek03a,lopez03}.
However, in some peculiar situations the effect remains almost
unchanged. Suppression of the Kondo anomaly is a consequence of an
effective exchange field due to coupling between the dot and
ferromagnetic electrodes. The exchange field gives rise to
spin-splitting of the equilibrium Kondo peak in DOS, and the two
components of the splitted peak move away from the Fermi level,
which leads to suppression of the Kondo anomaly in electrical
conductance -- similarly as an external magnetic field suppresses
the effect in nonmagnetic systems. Such a suppression was studied
recently by the Green function technique in the limit of infinite
correlation parameter $U$ \cite{martinek03a}, and was also
confirmed by numerical renormalization group calculations
\cite{choi03,martinek03b}. However, only QDs symmetrically coupled
to two magnetic leads were studied up to now. The Kondo anomaly
survives then in the antiparallel magnetic configuration and is
significantly suppressed in the parallel one. Recent experimental
observations on $C_{60}$ molecules attached to ferromagnetic (Ni)
electrodes support these general theoretical predictions
\cite{pasupathy04}.

Some features of the non-equilibrium Kondo phenomenon in QDs
coupled to ferromagnetic leads have not been addressed yet.
Therefore, in this paper we consider a more general situation.
First of all, we consider the case when the two ferromagnetic
electrodes are generally different. In other words, the dot is
(spin-)asymmetrically coupled to the ferromagnetic leads. This
leads to qualitatively new results. Second, we consider the case
of arbitrary $U$ instead of the limiting situation of infinite $U$
studied in [\onlinecite{martinek03a}]. Third, we introduce an
effective exchange field to describe the dot level
renormalization.

We analyze in detail three different situations. In the first case
the dot is coupled to two ferromagnetic leads, and the coupling is
fully symmetric in the parallel magnetic configuration. We show
that the equilibrium Kondo peak in DOS is then spin-split in the
parallel configuration, whereas no splitting appears in the
antiparallel one. The splitting, however, is significantly reduced
for small values of the correlation parameter $U$. The
corresponding zero-bias anomaly in conductance becomes split in
the parallel configuration as well \cite{martinek03a}. The second
situation studied in this paper is the one with asymmetric
coupling to two ferromagnetic leads. As a particular case we
consider the situation when one of the ferromagnetic electrodes is
half metallic, with almost total spin polarization of electron
states at the Fermi level. Such structures have been shown
recently to have transport characteristics with typical diode-like
behavior \cite{rudzinski01,swirkowicz03b}. Finally, we also
analyze the case when one of the electrodes is nonmagnetic whereas
the second one is ferromagnetic, and show that one ferromagnetic
electrode is sufficient to generate spin-splitting of the Kondo
anomaly.

The paper is organized as follows. The model and method are
briefly described in Sections 2 and 3, respectively. Numerical
results for the three different situations mentioned above are
presented and discussed in Section 4. Summary and general
conclusions are given in Section 5.

\section{model}

We consider a single-level QD coupled to ferromagnetic metallic
leads (electrodes) by tunnelling barriers. We restrict our
considerations to collinear (parallel and antiparallel) magnetic
configurations and assume that the axis z (spin quantization axis)
points in the direction of the net spin of the left electrode
(opposite to the corresponding magnetic moment). Antiparallel
alignment is obtained by reversing magnetic moment of the right
electrode. The whole system is then described by Hamiltonian of
the general form
\begin{equation}
H=H_{L}+H_{R}+H_{D}+H_{T}.
\end{equation}
The terms $H_{\beta}$ describe here the left ($\beta=L$) and right
($\beta=R$) electrodes in the non-interacting quasi-particle
approximation,
$H_{\beta}=\sum_{k\sigma}\varepsilon_{k\beta\sigma}c_{k\beta\sigma}
^{+}c_{k\beta\sigma}$, where $\varepsilon_{k\beta\sigma}$ is the
single-electron energy in the electrode $\beta$ for the
wave-number k and spin $\sigma$ ($\sigma =\uparrow$,
$\downarrow$), whereas $c_{k\beta\sigma}^{+}$ and
$c_{k\beta\sigma}$ are the corresponding creation and annihilation
operators. The single-particle energy $\varepsilon_{k\beta\sigma}$
includes the electrostatic energy due to applied voltage,
$\varepsilon_{k\beta\sigma}=
\varepsilon_{k\beta\sigma}^{0}+eU_{e}^{\beta}=\varepsilon_{k\beta\sigma}^{0}+\mu_{\beta}$,
where $\varepsilon_{k\beta \sigma}^{0}$ is the corresponding
energy in the unbiased system, $U_{e}^{\beta}$ is the
electrostatic potential of the $\beta$-th electrode, $e$ stands
for the electron charge ($e<0$), and $\mu_{\beta}$ is the chemical
potential of the $\beta$-th electrode (the energy is measured from
the Fermi level of unbiased system). Electron spin projection on
the global quantization axis is denoted as $\uparrow$ for
$s_z=1/2$ and $\downarrow$ and for $s_z=-1/2$. On the other hand,
spin projection on the local quantization axis (local spin
polarization in the ferromagnetic material) will be denoted as $+$
for spin-majority and $-$ for spin-minority electrons,
respectively. When local and global quantization axes coincide,
then $\uparrow$ is equivalent to $+$ and $\downarrow$ is
equivalent to $-$. (Note, that the local quantization axis in the
ferromagnet is opposite to the local magnetization.)

The term $H_{D}$ in Eq.(1) describes the quantum dot and takes the
form
\begin{equation}
H_{D}=\sum_{\sigma}\epsilon_\sigma\,
d_{\sigma}^{+}d_{\sigma}+Ud_{\uparrow}
^{+}d_{\uparrow}d^{+}_{\downarrow}d_{\downarrow}\, ,
\end{equation}
where $\epsilon_\sigma$ denotes energy of the dot level
(spin-dependent in a general case), $U$ denotes the electron
correlation parameter, whereas $d_{\sigma}^{+}$ and $d_{\sigma}$
are the creation and annihilation operators for electrons on the
dot. The level energy $\epsilon_\sigma$ includes the electrostatic
energy due to applied voltage, $\epsilon_\sigma
=\epsilon_{0\sigma}+eU_{e}^d$, where $U_{e}^d$ is the
electrostatic potential of the dot, and $\epsilon_{0\sigma}$ is
the level energy at zero bias.

The electrostatic potential $U_{e}^d$ of the dot will be
determined fully self-consistently from the following capacitance
model \cite{wang99,swirkowicz02}:
\begin{equation}
e\left(\,\sum_{\sigma}n_{\sigma}-\sum_{\sigma}n_{0\sigma}
\right)=C_{L}(U_{e}^d-U_{e} ^{L})+C_{R}(U_{e}^d-U_{e}^{R}),
\end{equation}
where $n_{\sigma}$ and $n_{0\sigma}$ are the dot occupation
numbers $\langle d^+_\sigma d_\sigma\rangle$ calculated for a
given bias and for zero bias, respectively, whereas $C_{L}$ and
$C_{R}$ denote the capacitances of the left and right tunnel
junctions. Self-consistent determination of the dot electrostatic
potential makes the description gauge invariant. This is
particularly important for strongly asymmetric systems.

The last term, $H_{T}$, in Eq.(1) describes tunnelling processes
between the dot and electrodes and is of the form
\begin{equation}
H_{T}=\sum_{k\beta\sigma}V_{k\beta\sigma}^{\ast}c_{k\beta\sigma}^{+}d_{\sigma
}+{\rm h.c.}\, ,
\end{equation}
where $V_{k\beta\sigma}$ are the components of the tunnelling
matrix, and ${\rm h.c.}$ stands for the Hermitian conjugate term.
Hamiltonian (4) includes only spin-conserving tunnelling
processes.

\section{Theoretical formulation}

Electric current flowing from the $\beta$-th lead to the quantum
dot in a nonequilibrium situation is determined by the retarded
(advanced) $G_{\sigma }^{r(a)}$ and correlation (lesser)
$G_{\sigma }^{<}$ Green functions of the dot (calculated in the
presence of coupling to the electrodes), and is given by the
formula \cite{jauho94}
\begin{equation}
I^\beta_\sigma=\frac{ie}{\hbar} \int
\frac{dE}{2\pi}\Gamma_\sigma^\beta (E)\{
G_{\sigma}^{<}(E)+f_{\beta}(E)
[G_{\sigma}^{r}(E)-G_{\sigma}^{a}(E)]\},
\end{equation}
where $f_{\beta}(E)$ is the Fermi distribution function for the
$\beta$-th electrode. The retarded (advanced) Green functions can
be calculated from the corresponding equation of motion. The key
difficulty is with calculating the lesser Green function
$G_{\sigma}^{<}(E)$.

In a recent paper \cite{swirkowicz03a} we applied the equation of
motion method to derive both $G_{\sigma}^{r}(E)$ and
$G_{\sigma}^{<}(E)$ Green functions within the same approximation
scheme \cite{niu99}. However, the approximations for the lesser
Green function $G_{\sigma}^{<}(E)$ do not conserve charge current
in asymmetrical systems. Therefore, in the following we assume
constant (independent of energy) coupling parameters,
$\Gamma^{\beta}_\sigma(E)=2\pi
\sum_{k}V_{k\beta\sigma}V_{k\beta\sigma}^{\ast}\delta
(E-\varepsilon_{k\beta\sigma})=\Gamma^{\beta}_\sigma$. As pointed
out in Refs [\onlinecite{kang,sun}], it is then sufficient to
determine $\int (dE/2\pi )\,G_{\sigma}^{<}(E)$, while knowledge of
the exact form of $G_{\sigma}^{<}(E)$ is not necessary. Current
conservation condition allows then to express the above integral
by an integral including retarded and advanced Green functions
only, which in turn allows to rewrite the current formula in the
commonly used form,
\begin{equation}
I_{\sigma}=\frac{ie}{\hbar} \int
\frac{dE}{2\pi}\frac{\Gamma_{\sigma}^{L}\Gamma_{\sigma}^{R}
}{\Gamma _{\sigma}^L+\Gamma _{\sigma}^R}
[G_{\sigma}^{r}(E)-G_{\sigma}^{a}(E)][f_{L}(E)-f_{R}(E)].
\end{equation}
Similarly, the occupation numbers, $n_{\sigma }=\langle d_{\sigma
}^{+}d_{\sigma }\rangle $, are then given by the formula
\[
n_{\sigma}=-i\int\frac{dE}{ 2\pi}\,G_{\sigma}^{<}(E)
\]
\begin{equation}
=i\int\frac{dE}{
2\pi}\frac{\Gamma_{\sigma}^{L}f_{L}(E)+\Gamma_{\sigma}
^{R}f_{R}(E)}{\Gamma_{\sigma}^L+\Gamma
_{\sigma}^R}[G_{\sigma}^{r}(E)-G_{\sigma } ^{a}(E)].
\end{equation}

In the following the parameters $\Gamma_{\sigma}^\beta$ will be
used to parameterize strength of the coupling between the dot and
leads. It is convenient to introduce the spin polarization factors
$p_\beta$ defined as $p_\beta =(\Gamma_+^\beta
-\Gamma_-^\beta)/(\Gamma_+^\beta +\Gamma_-^\beta)$, where
$\Gamma_+^\beta$ and $\Gamma_-^\beta$ are the coupling parameters
for spin-majority and spin-minority electrons in the lead $\beta$,
respectively. Accordingly, one may write $\Gamma_+^\beta
=(1+p_\beta )\Gamma^\beta$ and $\Gamma_-^\beta =(1-p_\beta
)\Gamma^\beta$, with $\Gamma^\beta =(\Gamma_+^\beta
+\Gamma_-^\beta )/2$.

The retarded (advanced) Green function $G_{\sigma}^{r(a)}$ of the
dot can be calculated only approximately, for instance by the
equation of motion method. In the approximations introduced by
Meir et al {\cite{meir93} one finds
\begin{equation}
G_{\sigma }^{r}(E)=\frac{E-\epsilon_\sigma -U(1-n_{-\sigma
})}{[E-\epsilon_\sigma -\Sigma _{\sigma
}^{r}(E)](E-\epsilon_\sigma -U)-Un_{-\sigma }\Sigma _{\sigma
}^{r}(E)},
\end{equation}
where $\Sigma _{\sigma }^{r}$ is the corresponding self energy,
\begin{equation}
\Sigma _{\sigma }^{r}(E)=\Sigma _{0\sigma
}^{r}(E)+U\frac{(E-\epsilon_\sigma )n_{-\sigma }\Sigma _{03\sigma
}^{r}(E)-L_{0\sigma }\Sigma _{01\sigma }^{r}(E)}{ L_{0\sigma
}(E)[L_{0\sigma }(E)-\Sigma _{03\sigma }^{r}(E)]},
\end{equation}
with $L_{0\sigma }=E-\epsilon_\sigma -U(1-n_{-\sigma })$, and
$\Sigma _{01\sigma }^{r }(E)$ and $ \Sigma _{03\sigma }^{r }(E)$
defined as
\begin{equation}
\Sigma _{01\sigma }^{r}(E)=n_{-\sigma }\Sigma _{0\sigma }^{r
}(E)+\Sigma _{1\sigma }^{r}(E),
\end{equation}
\begin{equation}
\Sigma _{03\sigma }^{r}(E)=\Sigma _{0\sigma }^{r}(E)+\Sigma
_{3\sigma }^{r}(E).
\end{equation}
The self energies $\Sigma _{0\sigma }^{r}(E)$, $\Sigma _{1\sigma
}^{r}(E)$, and $\Sigma _{3\sigma }^{r}(E)$ are defined as
\[
\Sigma _{0\sigma }^{r}(E)=\sum_{\beta =L,R}\Sigma _{0\sigma
}^{\beta r }(E)=\sum_{\beta =L,R}\sum_{k}\left| V_{k\beta \sigma
}\right| ^{2}\frac{1}{E-\varepsilon_{k\beta\sigma} +i0^+}
\]
\begin{equation}
= \sum_{\beta =LR}\int \frac{d\varepsilon }{2\pi }\frac{\Gamma
_{\sigma }^{\beta }}{E-\varepsilon} -i\frac{\Gamma _{\sigma
}^{\beta }}{2}
\end{equation}
and
\[
\Sigma _{i\sigma }^r(E)=\sum_{\beta =LR}\int \frac{d\varepsilon
}{2\pi }A_{i}\Gamma _{-\sigma }^{\beta
}\biggl[-\frac{1}{\varepsilon
-E-\epsilon_{-\sigma}+\epsilon_{\sigma}-i\hbar /\tau_{-\sigma} }
\]
\begin{equation}
+\frac{1}{\varepsilon +E-\epsilon_{-\sigma}
-\epsilon_{\sigma}-U-i\hbar /\tau_{-\sigma} } \biggr],
\end{equation}
(for i=1,3), where $A_1=f(\varepsilon)$, $A_3=1$, and
$\tau_{-\sigma}$ is the relaxation time of the intermediate states
\cite{meir93}. This relaxation time is spin dependent and in the
low-temperature limit is given by the formula \cite{meir93}
\[
\frac{1}{\tau_\sigma }  = \frac{1}{2\pi\hbar}\; \sum_{\beta
,\beta^\prime } \; \sum_{\sigma^\prime}\, \Gamma_\sigma^\beta \,
\Gamma_{\sigma^\prime}^{\beta^\prime} \, \Theta
(\mu_{\beta^\prime} -\mu_\beta +\epsilon_{\sigma}
-\epsilon_{\sigma^\prime} )
\]
\begin{equation}
\times \frac{\mu_{\beta^\prime} -\mu_\beta +\epsilon_{\sigma}
-\epsilon_{\sigma^\prime}}{(\mu_\beta
-\epsilon_{\sigma})(\mu_{\beta^\prime}
-\epsilon_{\sigma^\prime})},
\end{equation}
where $\Theta (x)=0$ for $x<0$ and $\Theta (x)=1$ otherwise.

In the limit of infinite $U$ the Green function (8) reduces to the
well known form,
\begin{equation}
G^r_\sigma (E)=\frac{1-n_{-\sigma}}{E-\epsilon_{\sigma}
-\Sigma^r_{0\sigma}(E)-\Sigma^r_{1\sigma}(E)},
\end{equation}
where $\Sigma _{0\sigma }^{r}(E)$  is given by Eq.(12) and
$\Sigma^r_{1\sigma}(E)$ by
\[
\Sigma _{1\sigma }^{r}(E) =\sum_{\beta
=L,R}\Sigma_{1\sigma}^{\beta r}(E)
\]
\begin{equation}
\equiv \sum_{\beta =L,R} \int \frac{d\varepsilon }{2\pi }\frac
{f_\beta (\varepsilon )\Gamma _{-\sigma }^\beta}{-\varepsilon
+E+\epsilon_{-\sigma}-\epsilon_{\sigma}+i\hbar /\tau_{-\sigma} },
\end{equation}
and $\tau_{\sigma}$ defined by Eq.(14).

The above derived Green functions are sufficient to describe
qualitatively basic features of the Kondo phenomenon in QDs
attached to nonmagnetic leads \cite{meir93}. However, they are not
sufficient to describe properly the Kondo phenomenon when the
quantum dot is attached to ferromagnetic leads. The key feature of
the system, which is not sufficiently taken into account is the
splitting of the dot level due to spin dependent tunneling
processes \cite{martinek03a}. The most natural way would rely on
an extension of the Green function calculations by going beyond
the approximations used to derive Eq.(8). This, however, leads to
cumbersome expressions. To avoid this, one may treat the problem
approximately by introducing 'by hand' the level splitting to the
formalism described above. One way to achieve such an objective
was proposed in Ref.\cite{martinek03a} for the limit of infinite
$U$, where the bare dot level $\epsilon_\sigma$ in
$\Sigma^r_{1\sigma}(E)$ and $\Sigma^r_{0\sigma}(E)$ was replaced
by the renormalized energy $\widetilde{\epsilon}_\sigma$
calculated in a self-consistent way. Such a renormalization works
well in the limit of infinite $U$. It is however not clear how to
extend it to the general case of arbitrary $U$. Therefore, we
decided to include the level splitting {\it via} an effective
exchange field. The exchange-induced spin-splitting of the level
is particularly important for the self-energies $\Sigma _{1\sigma
}^{r}(E)$ and $\Sigma _{3\sigma }^{r}(E)$, so we replace the bare
energy levels in the self-energies $\Sigma _{1\sigma }^{r}(E)$ and
$\Sigma _{3\sigma }^{r}(E)$ by the corresponding renormalized
levels $\widetilde{\epsilon}_\sigma  = \epsilon_\sigma \pm
g\mu_BB_{\rm ex}/2$, with the exchange field calculated from the
formula
\[
B_{\rm ex}=\frac{1}{g\mu_B}\sum_{\beta =R,L} {\rm Re}
\int\frac{d\varepsilon}{2\pi}f_\beta (\varepsilon)
\]
\[
\times \left[\Gamma^\beta_\uparrow \left( \frac{1}{\varepsilon
-\epsilon_{\uparrow} -i\hbar /\tau_\uparrow}- \frac{1}{\varepsilon
-\epsilon_{\uparrow} -U-i\hbar /\tau_\uparrow}\right) \right.
\]
\begin{equation}
\left. -\Gamma^\beta_\downarrow \left( \frac{1}{\varepsilon
-\epsilon_{\downarrow} -i\hbar /\tau_\downarrow}-
\frac{1}{\varepsilon -\epsilon_{\downarrow} -U-i\hbar
/\tau_\downarrow}\right)\right].
\end{equation}
When the bare dot level and spin relaxation time are independent
of the spin orientation, $\epsilon_\uparrow = \epsilon_\downarrow
= \epsilon $ and $\tau_\uparrow =\tau_\downarrow =\tau$, the
formula (14) acquires the form \cite{braun04}
\[
B_{\rm ex}=\frac{1}{g\mu_B}\sum_{\beta =R,L} {\rm Re}
\int\frac{d\varepsilon}{2\pi}f_\beta (\varepsilon)
\]
\begin{equation}
\times \left(\Gamma^\beta_\uparrow -\Gamma^\beta_\downarrow
\right)\left( \frac{1}{\varepsilon -\epsilon -i\hbar /\tau}-
\frac{1}{\varepsilon -\epsilon -U-i\hbar /\tau}\right) .
\end{equation}

To some extent such an approach is similar to that used in
Ref.[\onlinecite{martinek03a}], and both approaches give similar
results in the limit of large $U$. This follows from the fact that
the expression (17) for exchange field is basically the expression
for the self energy $\Sigma^r_{1\sigma}$ (see Eq.(13) for $i=1$
and $E=\epsilon_\sigma$). However, the approach based on the
exchange field allows to handle easily also the general case of
finite $U$.

\section{Numerical results}

Now, we apply the above described formalism to the Kondo problem
in a QD coupled to ferromagnetic leads. In the numerical
calculations described below the energy is measured in the units
of $D$, where $D=\bar{D}/50$ and $\bar{D}$ is the electron band
width. For simplicity, the electron band in the leads is assumed
to be independent of the spin orientation and extends from $-25D$
below the Fermi level (bottom band edge) up to $25D$ above the
Fermi level (top band edge). The energy integrals will be cut off
at $E=\pm 25D$, i.e., will be limited to the electron band. Thus,
the influence of ferromagnetic electrodes is included only {\it
via} the spin asymmetry of the coupling parameters
$\Gamma^L_\sigma$ and $\Gamma^R_\sigma$. Apart from this, in all
numerical calculations we assumed $kT/D=0.001$ and
$\epsilon_{0\uparrow}/D=\epsilon_{0\downarrow}/D=\epsilon_{0}/D=-0.35$.

For positive (negative) bias we assume the electrostatic potential
of the left (right) electrode equal to zero. In other words,
electrochemical potential of the drain electrode is assumed to be
zero while of the source electrode is shifted up by $\vert
eV\vert$. In the following the bias is described by the
corresponding electrostatic energy $eV$. Note that positive $eV$
corresponds to negative bias due to negative electron charge
($e<0$).

\subsection{QD coupled to two similar ferromagnetic leads}

Consider first the situation when both electrodes are made of the
same ferromagnetic metal, and the coupling of the dot to both
leads is symmetrical (in the parallel configuration). For
numerical calculations we assumed $\Gamma^L_+/D
=\Gamma^R_+/D=0.12$ for spin-majority electrons and $\Gamma^L_-/D
=\Gamma^R_-/D=0.08$ for spin-minority ones, which corresponds to
the spin polarization factor $p_L=p_R=p=0.2$, and
$\Gamma^L/D=\Gamma^R/D=0.1$.

\begin{figure}
\begin{center}
\includegraphics[width=0.7\columnwidth]{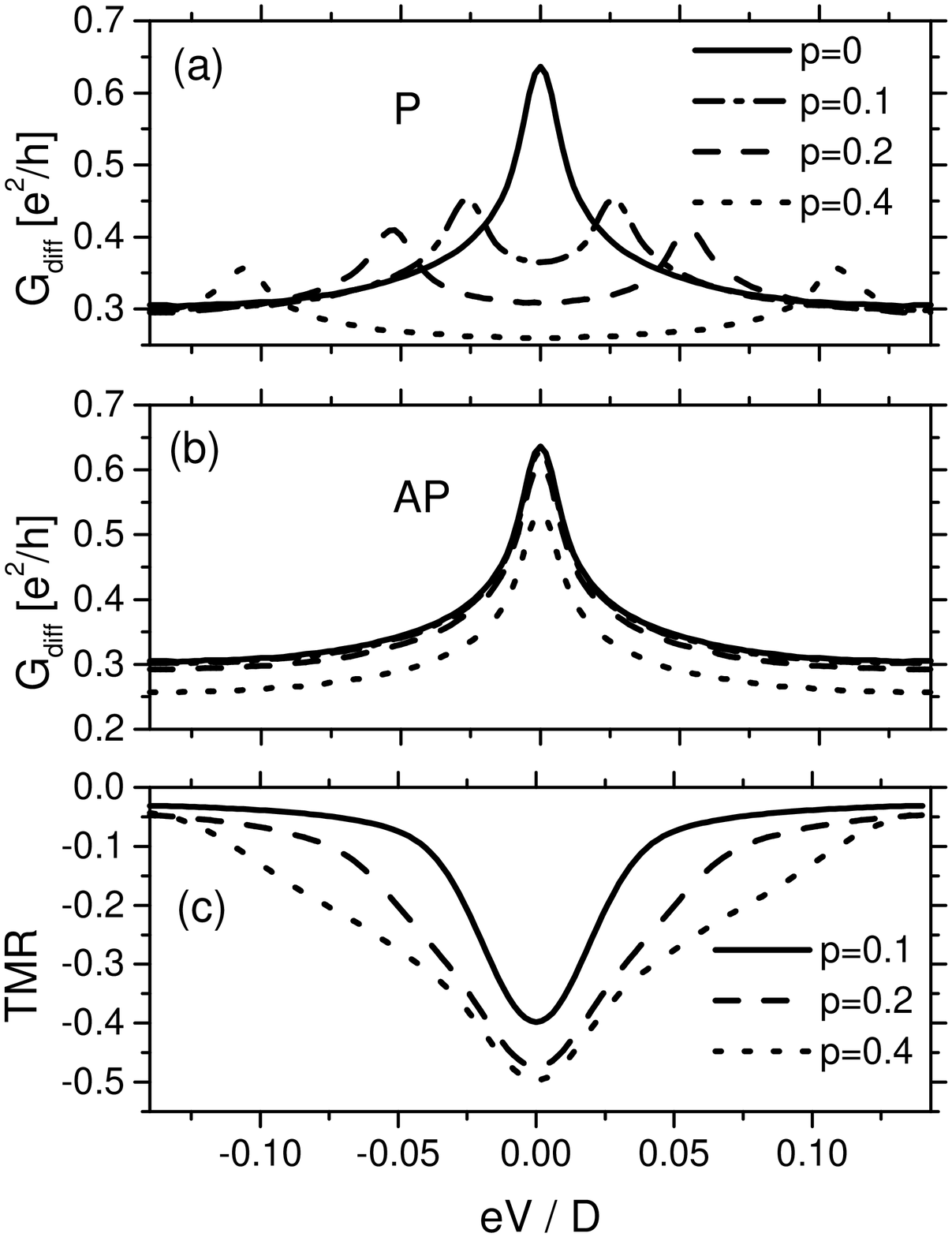}
\caption{Bias dependence of the differential conductance in the
parallel (a) and antiparallel (b) configurations, and the
corresponding TMR (c) for indicated four values of the lead
polarization $p_L=p_R=p$ ($p=0$ corresponds to nonmagnetic leads
so the corresponding TMR vanishes and is not shown in part (c)).
The parameters assumed for numerical calculations are:
$kT/D=0.001$, $\Gamma^L/D=\Gamma^R/D=0.1$,
$\epsilon_{0\uparrow}/D=\epsilon_{0\downarrow}/D=\epsilon_{0}/D=-0.35$,
$U/D=500$, and $(e^2/C_L)/D=(e^2/C_R)/D=0.33$.}
\end{center}
\end{figure}

It was shown in Ref.[\onlinecite{martinek03a}] that ferromagnetism
of the electrodes leads to spin splitting of the Kondo peak in the
density of states (DOS) in the parallel configuration, whereas no
splitting occurs for the antiparallel orientation. Such a behavior
of DOS has a significant influence on the transport properties.
First of all, the Kondo peak in DOS leads to zero bias anomaly in
the differential conductance $G_{diff}=\partial I /\partial V$.
This anomaly is particularly interesting in the parallel
configuration, where the spin splitting of the Kondo peak in DOS
leads to splitting of the differential conductance, as shown in
Fig.1(a) for four different values of the electrode spin
polarization factor $p$ and for large $U$. For $p=0$ there is only
a single peak centered at zero bias. When the polarization factor
becomes nonzero, the peak becomes split into two components
located symmetrically on both sides of the original peak, with the
corresponding intensities significantly suppressed. The splitting
of the Kondo anomaly increases with increasing $p$. Moreover,
height of the two components of the Kondo anomaly decreases with
increasing $p$. On the other hand, in the antiparallel
configuration there is no splitting of the Kondo peak in the
density of states and consequently also no splitting of the Kondo
anomaly in the differential conductance (see Fig.1(b)). For all
polarization values, the anomaly is similar to that in the case of
QDs coupled to nonmagnetic leads. However, intensity of the Kondo
anomaly decreases with increasing polarization. Difference between
conductance in the antiparallel and parallel configurations gives
rise to the TMR effect which may be described quantitatively by
the ratio  $(I^{\rm P}-I^{\rm AP})/I^{\rm AP}$, where $I^{\rm P}$
and $I^{\rm AP}$ denote the current flowing through the system in
the parallel and antiparallel configurations at the same bias,
respectively. The associated TMR effect is displayed in Fig.1(c).
One finds negative values of the TMR ratio, which is a consequence
of the spin-splitting of the the Kondo peak in the parallel
configuration and absence of such a splitting for antiparallel
alignment. It is worth to note, that in the absence of the Kondo
anomaly the TMR effect would be positive.

\begin{figure}
\begin{center}
\includegraphics[width=0.7\columnwidth]{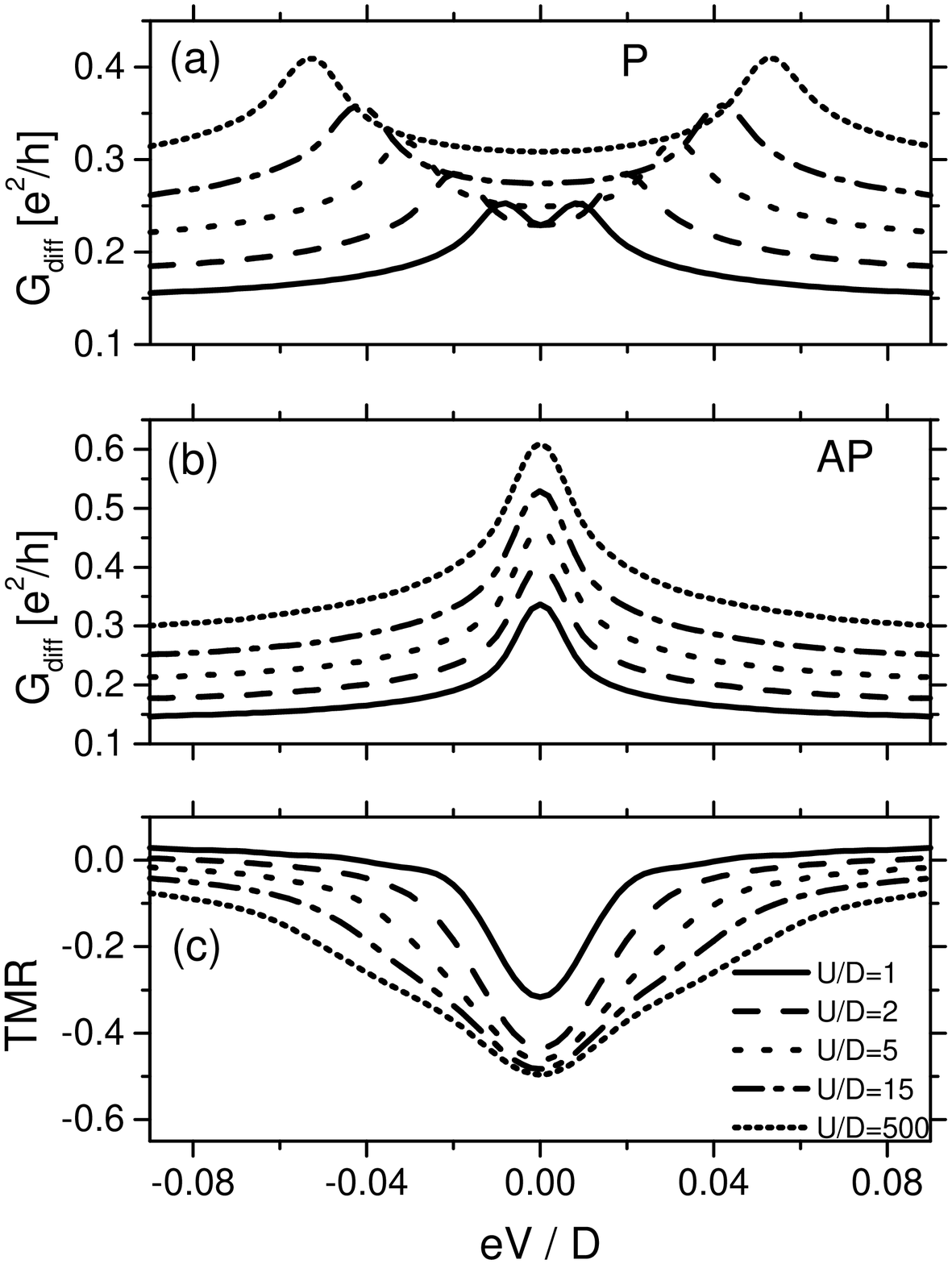}
\caption{Bias dependence of differential conductance in the
parallel (a) and antiparallel (b) configurations, and the
corresponding TMR (c) calculated for indicated values of the
correlation parameter $U$ and for $p=0.2$. The other parameters
are as in Fig.1.}
\end{center}
\end{figure}

The Kondo anomaly in transport characteristics shown in Fig.1 was
calculated for the limit of large $U$. In Fig.2 we show similar
characteristics as in Fig.1, but for different values of the
correlation parameter $U$ and a constant value of the polarization
factor $p$. The splitting of the Kondo anomaly in the parallel
configuration and intensity of the peaks (Fig.2(a)) decrease with
decreasing $U$. In the antiparallel configuration there is no
splitting of the Kondo anomaly, but intensity of the Kondo peak
decreases with decreasing $U$. The associated TMR effect is shown
in Fig.2(c). The effect is negative in a certain bias range around
the zero bias limit, but absolute magnitude of the effect becomes
smaller for smaller values of $U$. For large bias there is a
transition from negative to positive TMR with decreasing $U$.

\subsection{QD coupled to one ferromagnetic and one
half-metallic leads}

Assume now that one of the electrodes (say the left one) is made
of a 3-d ferromagnetic metal, the second (right) one is
half-metallic, and the total coupling to the latter electrode is
much smaller than to the former one. This is reflected in the spin
asymmetry of the bare coupling constants, for  which we assume
$\Gamma^L_+/D =0.28$ and $\Gamma^L_-/D =0.12$ for the left
electrode, and $\Gamma^R_+/D =0.04$ and $\Gamma^R_-/D =0.0002$ for
the right one. These parameters correspond to $p_L=0.4$,
$p_R=0.99$, $\Gamma^L/D=0.2$, and $\Gamma^R/D\approx 0.02$. Thus,
the spin asymmetry of the coupling to the right electrode is much
larger than to the left one.  In Fig.3 we show DOS in the parallel
(left column) and antiparallel (right column) magnetic
configurations, calculated for vanishing as well as for positive
and negative bias voltages. Consider now the main features of the
spectra in more details, and let us begin with the parallel
configuration (left column in Fig.3).

\begin{figure}
\begin{center}
\includegraphics[width=0.8\columnwidth]{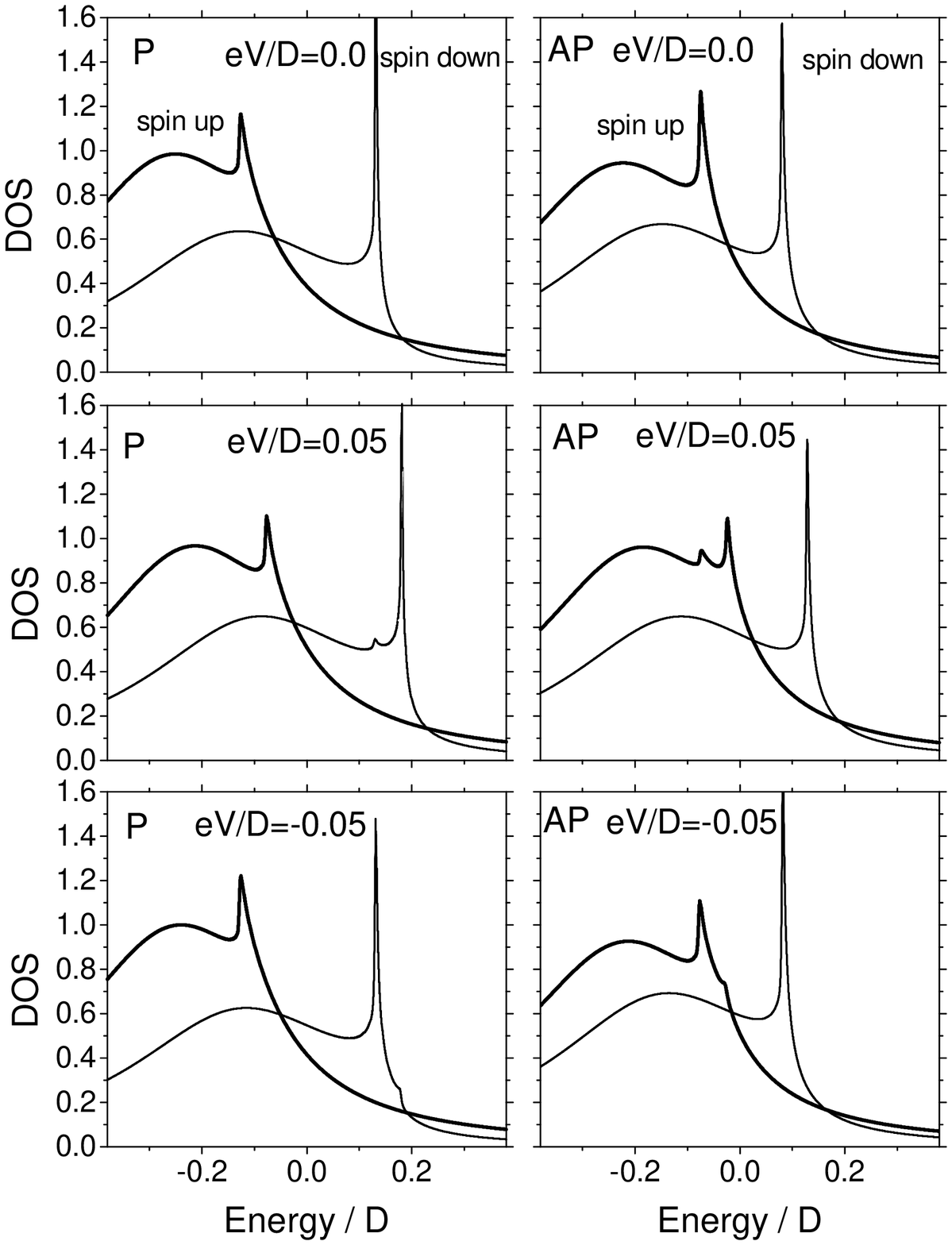}
\caption{DOS for spin-up (thick lines) and spin-down (thin lines)
electron states on the dot in the parallel (left column) and
antiparallel (right column) magnetic configurations, calculated
for three indicated voltages and for $\Gamma^L_+/D =0.28$,
$\Gamma^L_-/D =0.12$, $\Gamma^R_+/D=0.04$, $\Gamma^R_-/D=0.0002$,
$U/D=500$, $(e^2/C_L)/D=(e^2/C_R)/D=10$. The other parameters are
as in Fig.1.}
\end{center}
\end{figure}

At $V=0$ the Kondo peak in DOS is spin-split, and the intensity of
spin-down peak is relatively large, whereas that of the spin-up
peak is much smaller. The asymmetry in peak intensities is a
consequence of the spin asymmetry in the coupling of the dot to
metallic electrodes -- this coupling is larger for spin-up
electron, which determines hight of the Kondo peak for spin-down
electrons. When a bias voltage is applied, each of the two Kondo
peaks generally becomes additionally split into two components.
One of them (the one associated with the coupling to the source
electrode) moves up in energy, whereas position of the second one
(the one associated with the drain electrode) remains unchanged.
This is because we assumed that the electrochemical potential of
the source electrode shifts up by $\vert eV\vert$, while of the
drain electrode is independent of the voltage. For $eV>0$
(negative bias), the splitting of the large-intensity (spin-down)
peak is clearly visible, although one component of the splitted
peak is relatively small. This is just the component associated
with the coupling of the dot to the right electrode in the spin-up
channel. Since this coupling is relatively small, the
corresponding intensity is also small. The second component, in
turn, is much larger because it is associated with the coupling to
the left electrode in the spin-up channel, which is the dominant
coupling in the system considered. Splitting of the low-intensity
(spin-up) peak is not resolved. Intensity of the component
associated with the coupling to the right electrode in the
spin-down channel practically vanishes because this coupling is
negligible in the case considered. For $eV<0$ (positive bias), the
situation is changed. Now the electrochemical potential of the
left electrode is independent of the bias. Consequently, intensity
of the components whose position is independent of energy is
significantly larger than intensity of the other components (the
ones associated with the right electrode). As before, the
component associated with the coupling to the right electrode in
the spin-down channel is not resolved.

\begin{figure}
\begin{center}
\includegraphics[width=0.7\columnwidth]{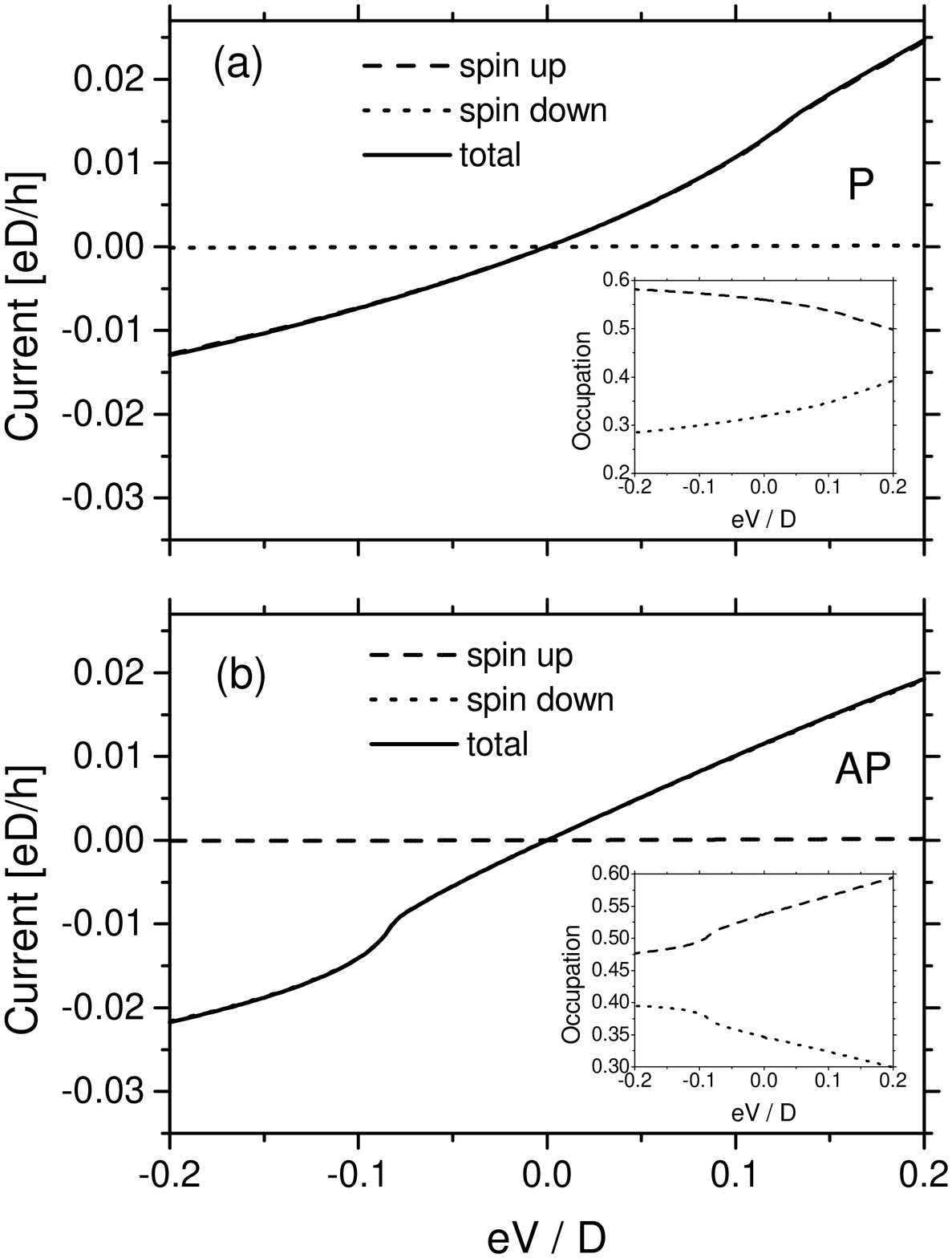}
\caption{Current-voltage characteristics in the parallel (a) and
antiparallel (b) configurations, calculated for the parameters as
in Fig.3. Current in the spin-down (spin-up) channel in the
parallel (antiparallel) configuration almost vanishes so the total
current flows in the spin-up (spin-down) channel (the curves
presenting the total and spin-up (spin-down) currents are not
resolved. The insets show the corresponding occupation numbers.
The parameters as in Fig.3}
\end{center}
\end{figure}

Consider now the antiparallel configuration (right column in
Fig.3), when magnetic moment of the right electrode is reversed.
There is a nonzero spin splitting of the Kondo peak at
equilibrium, contrary to the case of symmetric coupling to the
magnetic electrodes, where the spin splitting in the antiparallel
configuration vanishes \cite{martinek03a}. Apart from this, the
situation is qualitatively similar to the one for parallel
configuration. The main difference is that now the bias-induced
splitting of the large-intensity peak is not resolved, whereas
splitting of the low-intensity peak is resolved.

As in the case of symmetrical coupling described above, the Kondo
peaks in DOS give rise to anomalous behavior of the corresponding
transport characteristics. Due to the splitting of the equilibrium
Kondo peak, the anomaly in DOS does not contribute to transport in
the small bias regime. The Kondo peaks enter the 'tunnelling
window' at a certain bias, which leads to an enhanced conductance.
Such an enhancement is clearly visible in the current-voltage
characteristics shown in Fig.4 for both parallel (a) and
antiparallel (b) configurations (solid lines), where for negative
values of $eV$ the enhancement is quite significant, but it is
less pronounced for $eV>0$. This asymmetry is due to the
difference in intensities of the corresponding Kondo peaks that
enter the 'tunnelling window'.

\begin{figure}
\begin{center}
\includegraphics[width=0.7\columnwidth]{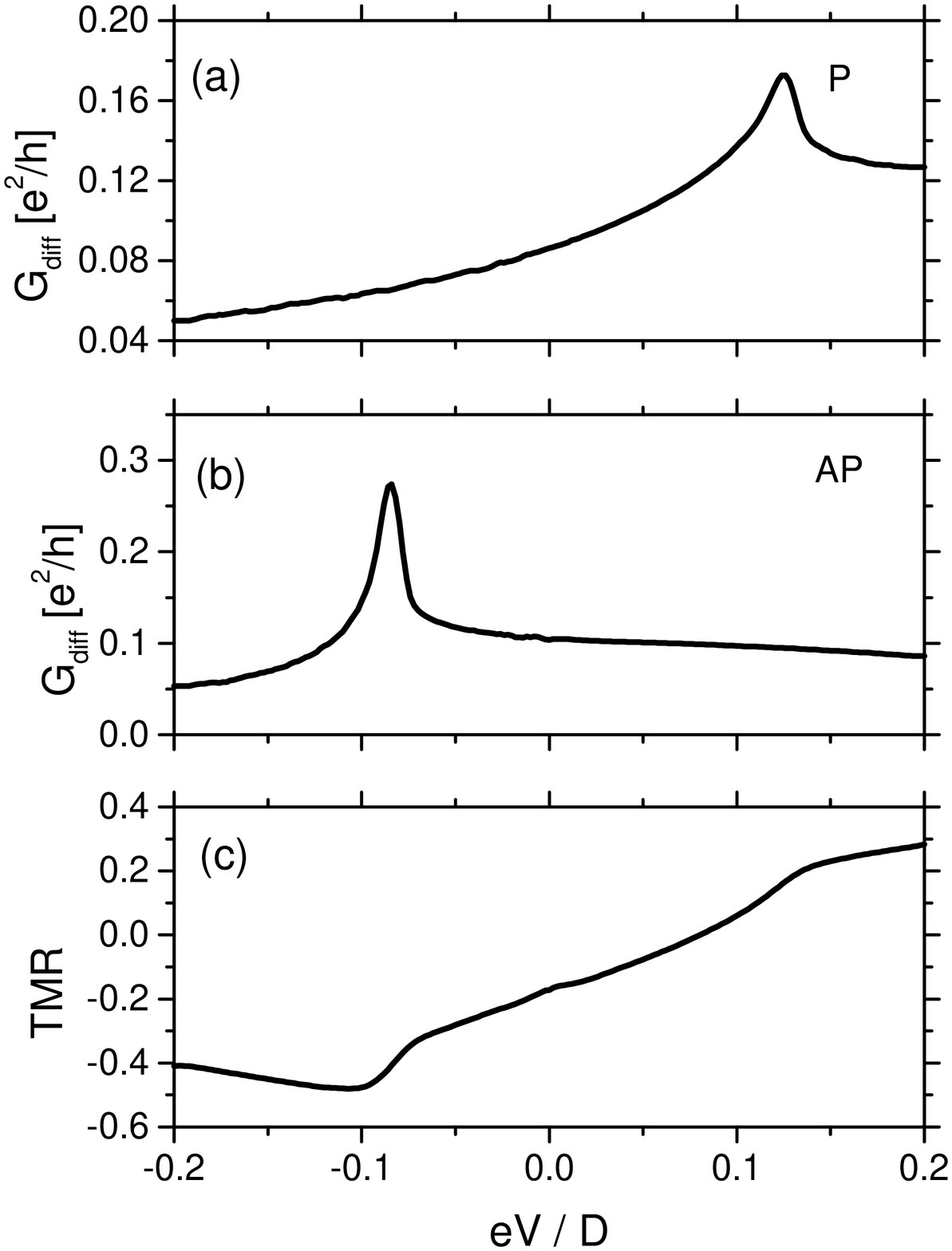}
\caption{Bias dependence of the differential conductance in the
parallel (a) and antiparallel (b) configurations and the
corresponding TMR (c), calculated for the same parameters as in
Fig.3.}
\end{center}
\end{figure}

The differential conductance in the Kondo regime is shown in Fig.5
for parallel (a) and antiparallel (b) magnetic configurations. In
the parallel configuration the Kondo anomaly occurs in the spin-up
channel and for $eV>0$ only. This may be easily understood by
considering the relevant DOS (see Fig.3, left column). For $eV>0$
only the Kondo peak in spin-up DOS can enter the tunnelling window
created by the bias. For $eV<0$, on the other hand, the Kondo peak
in spin-down DOS can enter the tunneling window. However, the spin
down channel is almost non-conducting, so the corresponding peak
in the differential conductance is suppressed. In the antiparallel
configuration the Kondo peak in differential conductance occurs
for $eV<0$ only. This can be accounted for by taking into account
behavior of the Kondo peaks in DOS shown in Fig.3 (right column),
and the fact that now the spin-up channel is non-conducting. For
$eV>0$ only the Kondo peak in the spin-up DOS can enter the
tunneling window, whereas for $eV<0$ this is the Kondo peak in
spin-down DOS (of large intensity).

The corresponding TMR is shown in Fig.5(c). It is interesting to
note that TMR is highly asymmetrical with respect to the bias
reversal. It becomes positive for $eV$ exceeding a certain
positive value, and negative below this voltage. This is a
consequence of the fact that for positive $eV$ the Kondo peak in
differential conductance is clearly visible in the parallel
configuration (see Fig.5(a)), whereas for $eV<0$ the Kondo peak
occurs in the antiparallel configuration (see Fig.5(b)). Such a
behavior of the conductance and also TMR may be interesting from
the point of view of applications in mesoscopic diodes.

\subsection{QD coupled to one ferromagnetic and one nonmagnetic
leads}

A specific example of asymmetric systems is the case where one
electrode is ferromagnetic (typical ferromagnetic 3d metal)
whereas the second one is nonmagnetic. For numerical calculations
we assumed $\Gamma^L_+/D =0.12$, $\Gamma^L_-/D =0.08$ for the left
(magnetic) electrode, and $\Gamma^R_+/D=\Gamma^R_-/D=0.1$ for the
right (nonmagnetic) one, which corresponds to $p_L=0.2$, $p_R=0$,
and $\Gamma^L/D=\Gamma^R/D=0.1$. As in the other asymmetrical
situations studied in this paper, the equilibrium Kondo peak in
DOS becomes spin-split. When a bias voltage is applied, each
component becomes additionally split, as shown in Fig.6. Variation
of the spectra with bias voltage can be accounted for in a similar
way as in the case of the dot coupled asymmetrically to two
ferromagnetic electrodes. The only difference is that now all
components of the peaks are clearly resolved. This is because all
coupling constants are now of comparable magnitude.

\begin{figure}
\begin{center}
\includegraphics[width=0.7\columnwidth]{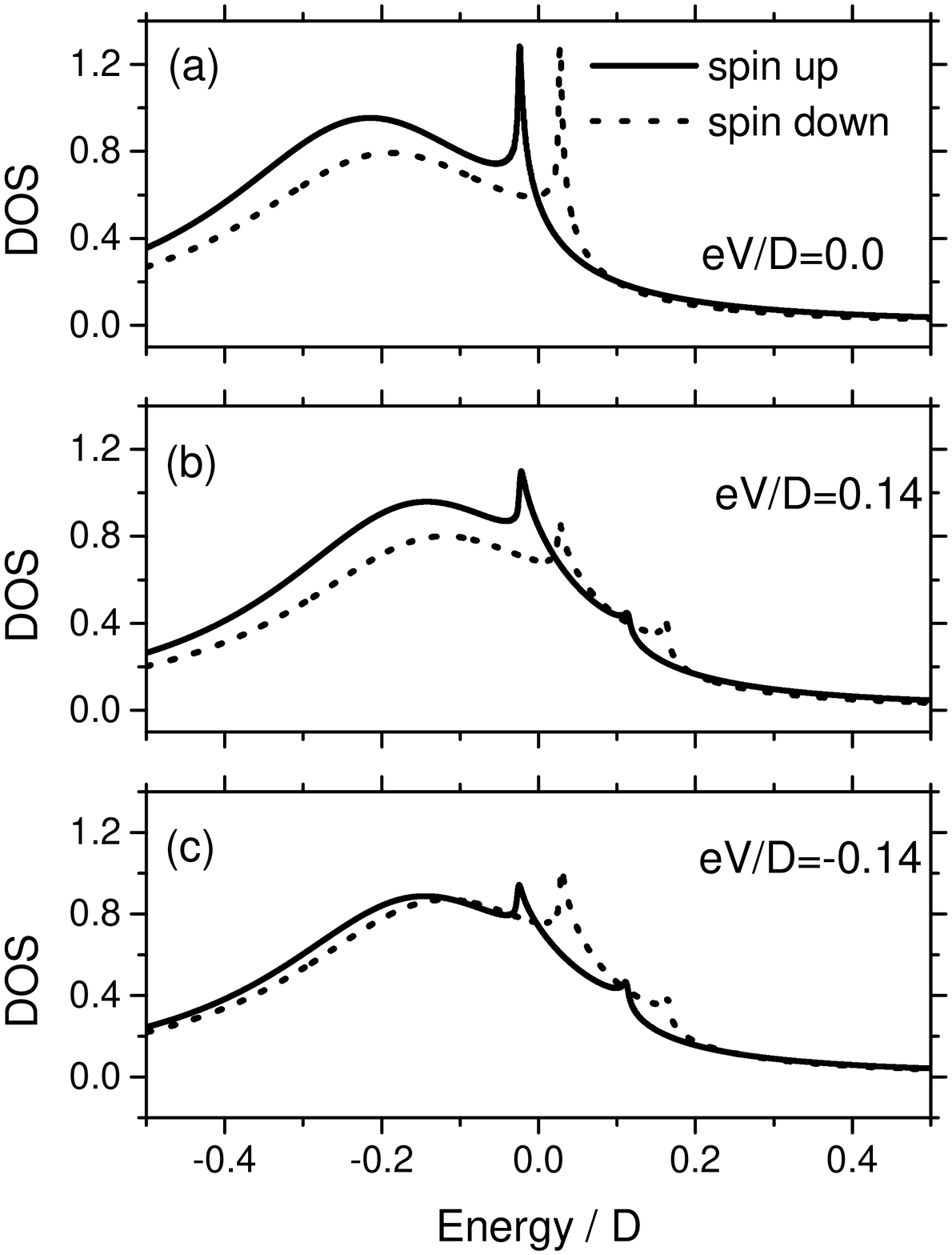}
\caption{DOS for spin-up (solid lines) and spin-down (dotted
lines) electron states of the dot, calculated for three different
voltages, and for $\Gamma^L_+/D =0.12$, $\Gamma^L_- /D=0.08$,
$\Gamma^R_+/D=\Gamma^R_-/D=0.1$, $U/D=500$, and
$(e^2/C_L)/D=(e^2/C_R)/D=0.33$. The other parameters are as in
Fig.1.}
\end{center}
\end{figure}

\begin{figure}
\begin{center}
\includegraphics[width=0.7\columnwidth]{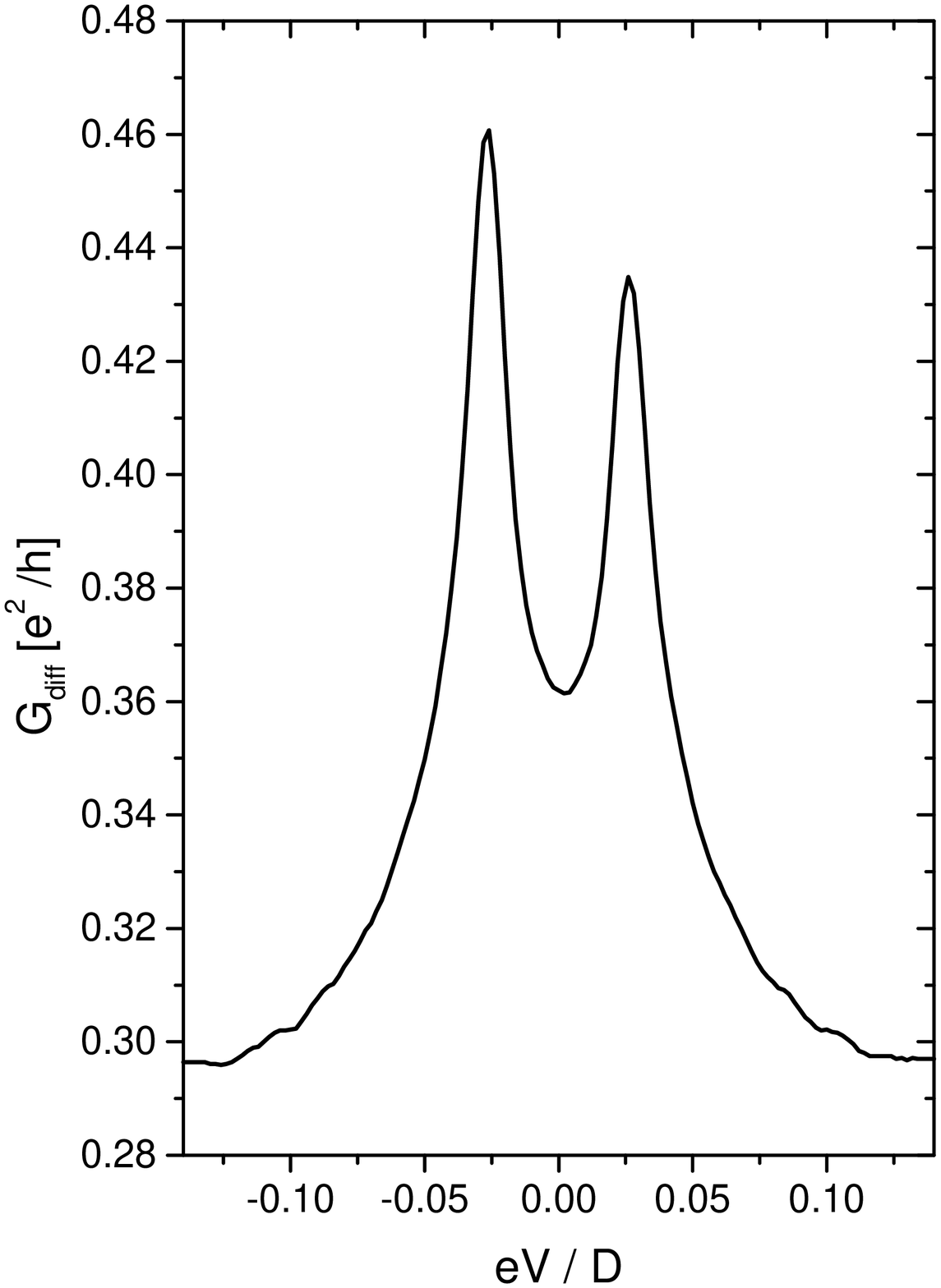}
\caption{Bias dependence of the differential conductance,
calculated for the  parameters as in Fig.6.}
\end{center}
\end{figure}

The corresponding differential conductance is shown in Fig.7. Due
to the spin splitting of the Kondo peak in DOS, the Kondo anomaly
in the conductance becomes split as well, as clearly seen in
Fig.7. However, the splitting is asymmetric with respect to the
bias reversal. Thus, there is no need to have two ferromagnetic
electrodes to observe splitting of the Kondo anomaly, but it is
sufficient when only one lead is ferromagnetic.

\section{Summary and conclusions}

In this paper we considered the Kondo problem in quantum dots
coupled symmetrically and asymmetrically to ferromagnetic leads.
As an specific example of asymmetrical systems, we considered the
case when one electrode is ferromagnetic, whereas the second one
is nonmagnetic.

We showed that ferromagnetism of the leads gives rise to a
splitting of the equilibrium Kondo peak in DOS for all
asymmetrical situations. This generally takes place for both
magnetic configurations when the two electrodes are different. The
splitting in both configurations also occurs when both magnetic
electrodes are of the same material, but the corresponding
coupling strengths to the dot are different. Indeed, such a
splitting in parallel and also antiparallel configurations was
recently observed experimentally \cite{pasupathy04}. When similar
electrodes are symmetrically coupled to the dot, the splitting
occurs only in the parallel configuration. An interesting
conclusion from the experimental point of view is that the
splitting also occurs in the case when one electrode is
nonmagnetic.

The spin-splitting of DOS can lead to characteristic splitting of
the zero bias anomaly in electrical conductance. This in turn can
lead to negative (inverse) tunnel magnetoresistance effect. In
highly asymmetrical systems TMR can change sign when bias voltage
is reversed.

\begin{acknowledgments}
The work was supported by the State Committee for Scientific
Research through the Research Project PBZ/KBN/044/P03/2001 and 4
T11F 014 24.
\end{acknowledgments}

\end{document}